\documentclass[times,authoryear]{elsarticle}

\usepackage{jasr}
\usepackage{framed,multirow}

\usepackage{amssymb}
\usepackage{latexsym}

\usepackage[switch]{lineno}

\usepackage{url}
\usepackage{xcolor}
\definecolor{newcolor}{rgb}{.8,.349,.1}

\usepackage[citebordercolor=white]{hyperref}

\journal{ASR}

\begin{document}

\verso{Chakraborty, S., \textit{et. al}}

\begin{frontmatter}

\title{Influence of ICME-driven Magnetic Cloud-like and Sheath Region induced Geomagnetic Storms in causing anomalous responses of the Low-latitude Ionosphere: A Case Study}%


\author[a,b]{Sumanjit Chakraborty\corref{c-d54cc1eb1ca4}}
\ead{sumanjit11@gmail.com}\cortext[c-d54cc1eb1ca4]{Corresponding author.}
\author[a]{Dibyendu Chakrabarty}
\ead{dipu@prl.res.in} 
\author[a]{Anil K. Yadav}
\ead{anil@prl.res.in}
\author[b]{Gopi K. Seemala}
\ead{gopi.seemala@gmail.com}

\affiliation[a]{organization={Space and Atmospheric Sciences Division},
                addressline={Physical Research Laboratory},
                postcode={Ahmedabad  380009},
                country={Gujarat, India}
                }
                
\affiliation[b]{organization={Indian Institute of Geomagnetism}, 
                postcode={Navi Mumbai 410218},
                country={Maharashtra, India}
                }

\received{xx}
\finalform{xx}
\accepted{xx}
\availableonline{xx}
\communicated{}

\begin{abstract}

This work shows an anomalously enhanced response of the low-latitude ionosphere over the Indian sector under weak geomagnetic conditions (October 31, 2021) in comparison to a stronger event (November 04, 2021) under the influence of an Interplanetary Coronal Mass Ejection (ICME)-driven Magnetic Cloud (MC)-like and sheath regions respectively. The investigation is based on measurements of the Total Electron Content (TEC) from Ahmedabad (23.06$^\circ$N, 72.54$^\circ$E, geographic; dip angle: 35.20$^\circ$), a location near the northern crest of the Equatorial Ionization Anomaly (EIA) over the Indian region. During the weaker event, the observed TEC from the Geostationary Earth Orbit (GEO) satellites of Navigation with Indian Constellation (NavIC), showed diurnal maximum enhancements of about 20 TECU over quiet-time variations, as compared to the stronger event where no such enhancements are present. It is shown that storm intensity (SYM-H) or magnitude of the southward Interplanetary Magnetic Field (IMF) alone is unable to determine the ionospheric impacts of this space weather event. However, it is the non-fluctuating southward IMF and the corresponding penetration electric fields, for a sufficient interval of time, in tandem with the poleward neutral wind variations, that determines the strengthening of low-latitude electrodynamics of this anomalous event of October 31, 2021. Therefore, the present investigation highlights a case for further investigations of the important roles played by non-fluctuating penetration electric fields in determining a higher response of the low-latitude ionosphere even if the geomagnetic storm intensities are significantly low. 

\end{abstract}


\end{frontmatter}


\section{Introduction}
\label{sec1}

Geomagnetic storms occur as a result of energy input from the solar wind into the Earth's Magnetosphere-Ionosphere (MI) system mostly following the Coronal Mass Ejection(s) (CMEs) from the Sun directed toward the Earth. These ICMEs have certain features \citep{sc:0,sc:00,sc:69} such as the sheath and the Magnetic Cloud (MC) regions. The sheath region, bounded by the shock front and formed ahead of the ICMEs, has the characteristic wherein the plasma is turbulent and highly compressed. The magnetic field, the proton temperature, proton density, dynamic flow pressure, and the velocity are all enhanced in an ICME sheath region. On the other hand, the MC region is characterized by a slow rotation of the magnetic field and much lower values of proton temperature \citep{sc:70,sc:71}.   

Geomagnetic disturbances begin when the north-south ($B_z$) component of the frozen-in Interplanetary Magnetic Field (IMF), carried by the solar wind, turns and remains southward, causing strong reconnection with the geomagnetic field \citep{sc:1,sc:1.1,sc:1.2,sc:2}. Following this phenomenon, the penetration of magnetospheric convection electric fields from the high latitudes to the low- and equatorial latitudes occurs which is known as the Prompt Penetration of Electric Field (PPEF) \citep{sc:3}. Studies have shown that the equivalent current system, associated with the PPEF, has a few minutes to about 3-4 hours of average lasting periods \citep{sc:4,sc:5}. The PPEF has been shown to have significant effects on the low-to-equatorial latitude electrodynamics \citep{sc:6,sc:7,sc:8}. Furthermore, at the poles and the higher latitudes, the Auroral Electrojet (AE) currents transfer heat energy to the neutral gas by the Joule heating process \citep{sc:36}. These currents, via the process of momentum transfer, cause changes in the circulation of the neutral wind in and around the sub-auroral thermosphere. As a result, the Disturbance Dynamo Electric Field (DDEF) plays a role in the equatorward transportation of plasma from the higher to the lower latitudes \citep{sc:37,sc:38,sc:39}. This DDEF is opposite (westward during daytime and eastward during nighttime) in nature to the PPEF and is a long-lasting process that lasts for a few days after the storm onset. During such periods, the DDEF modulates low-latitude electrodynamics significantly in comparison to the quiet-time variations.

The Equatorial Ionization Anomaly (EIA) characterizes the ionosphere over the low-latitude regions of the globe. The EIA gets developed at the northern and southern crests around $\pm15-20^\circ$ magnetic latitude mainly due to the removal of plasma from around the equator by the upward E $\times$ B drift and partly due to the diffusion of plasma along the field lines from higher altitudes \citep{sc:9,sc:10}. The daily variation of this EIA and hence the electron density or the Total Electron Content (TEC) over the low-latitude regions are due to the variations in the electric fields and the thermospheric winds \citep{sc:11}. 

Several researchers (\cite{sc:21,sc:19,sc:20,sc:22,sc:23,sc:24,sc:49} and references therein) have studied the response of the ionosphere to strong geomagnetic storms over different longitude sectors throughout the globe. However, there have been only a few studies that address the response of the ionosphere to weak-to-moderate geomagnetic storms, often ignored because of the expectation of negligible alterations in the ionospheric electrodynamics over daily variations. The study by \citep{sc:25} showed that geomagnetic storms of moderate intensity are capable of causing greater effects on the mid-latitude ionosphere while the study by \citep{sc:68} showed the observations of large ionospheric variations during a weak geomagnetic storm. A recent study by \citep{sc:26,sc:75.1} brought forward similar observations of anomalous enhancements in the TEC over the low-latitude regions of the Indian longitude sector. However, studies dedicated to low-latitude responses, around the EIA region, during geomagnetically weak as well as strong conditions, under the influence of an ICME-MC or sheath region, are sparse. 

The motivation for this work comes from the fact that whether a stronger IMF condition in terms of higher magnitude (due to different subsets of ICMEs) and subsequent enhancement in ring current (observed as Symmetric (SYM)-H decreases), would always necessarily mean a stronger response of the low-latitude ionosphere. This study shows the anomalous response of the ionosphere over a low-latitude region in the Indian sector under a weak geomagnetic storm (October 31, 2021) and compares the same under a stronger geomagnetic storm (November 04, 2021) due to  MC-like and ICME-driven sheath regions. The present investigation brings forward an anomalous case where for an ICME to be geoeffective, especially up to the point of perturbing the low-latitude ionosphere, it has to be the duration of the non-fluctuating or consistent southward IMF $B_z$, in addition to poleward conditions of the neutral wind, to determine the ionospheric responses over these regions, irrespective of the intensity of the geomagnetic storm. This case study becomes important for understanding changes in the low-latitude ionosphere during such cases of weak space weather events which often get ignored due to the general expectation of weaker impacts on the ionosphere under such conditions. The present study also becomes vital for developing a reliable space weather forecast system, in terms of incorporation, of the time duration effect of the southward IMF, into the upcoming state-of-the-art machine learning modules, over the low-latitude regions.   

The manuscript is divided as follows: Section 2 describes the data and the methodologies used in the present study. Section 3 describes the results, including the sources of geomagnetic storms, the interplanetary and SYM-H conditions, and the ionospheric conditions. Section 4 presents the discussions of the physical mechanisms to have produced such observations while Section 5 summarizes the work.  

\section{Data and Methodology}

The Indian Regional Navigation Satellite System (IRNSS), presently known as the Navigation with Indian Constellation (NavIC), is a regional navigation satellite system developed by the Indian Space Research Organization (ISRO). Three Geostationary Earth Orbit satellites (GEOs) and three Geo-Synchronous Orbit satellites (GSOs) comprise the space segment of NavIC which have continuous visibility on the ionosphere over the Indian region. These satellites broadcast in the L5 having a carrier frequency of 1176.45 MHz and in the S1 band having a carrier frequency of 2492.028 MHz. NavIC has been conceived primarily for the estimation of errors in positioning, mainly over the entire Indian region, that has the northern crest of EIA and the geomagnetic equator passing over it. Furthermore, the suitability of NavIC to study the upper atmosphere has been well-established by several researchers \citep{sc:55,sc:56,sc:57,sc:58,sc:59,sc:29.2,sc:75,sc:60,sc:62,sc:65,sc:66,sc:67,sc:90} as well as \cite{sc:29.1} where NavIC data are used to separate the spatial and temporal effects. It is important to note the consistency between the VTECs of NavIC and GPS, both under quiet and disturbed conditions and when compared with the various global ionospheric models over the Indian longitude sector \citep{sc:29.2,sc:29.3}, further solidifying its observational capabilities related to regional ionospheric research. 

The Physical Research Laboratory (PRL), Ahmedabad has been operating an Accord-make NavIC receiver, since July 2017, capable of receiving GPS L1 (1575.42 MHz) along with NavIC L5 and S1 signals. The Slant TEC ($STEC$) is one of the outputs given by the receivers. The $STEC$ is converted to the equivalent Vertical TEC ($VTEC$). This conversion (equation \ref{b}) uses a mapping function $M_f$ (equation \ref{a}) assuming the ionosphere as a thin shell at an altitude ($H$) of 350 km (\cite{sc:28,sc:73,sc:29} and references therein).

\begin{equation}
    M_f = \sqrt{1-\left(\frac{R_E*cos(\theta)}{R_E + H} \right)^2}
    \label{a}
\end{equation}

\begin{equation}
    VTEC = M_f^{-1}*STEC
    \label{b}
\end{equation}

Here $R_E$ = 6371 km is the radius of Earth and $\theta$ are the elevation angles of the satellites. It is to be noted that to reduce the effects of multipath error on the NavIC signals, elevation angles $>30^{\circ}$ have been chosen for the entire analysis. 

\begin{figure}[ht]
\centering
\includegraphics[scale=0.35]{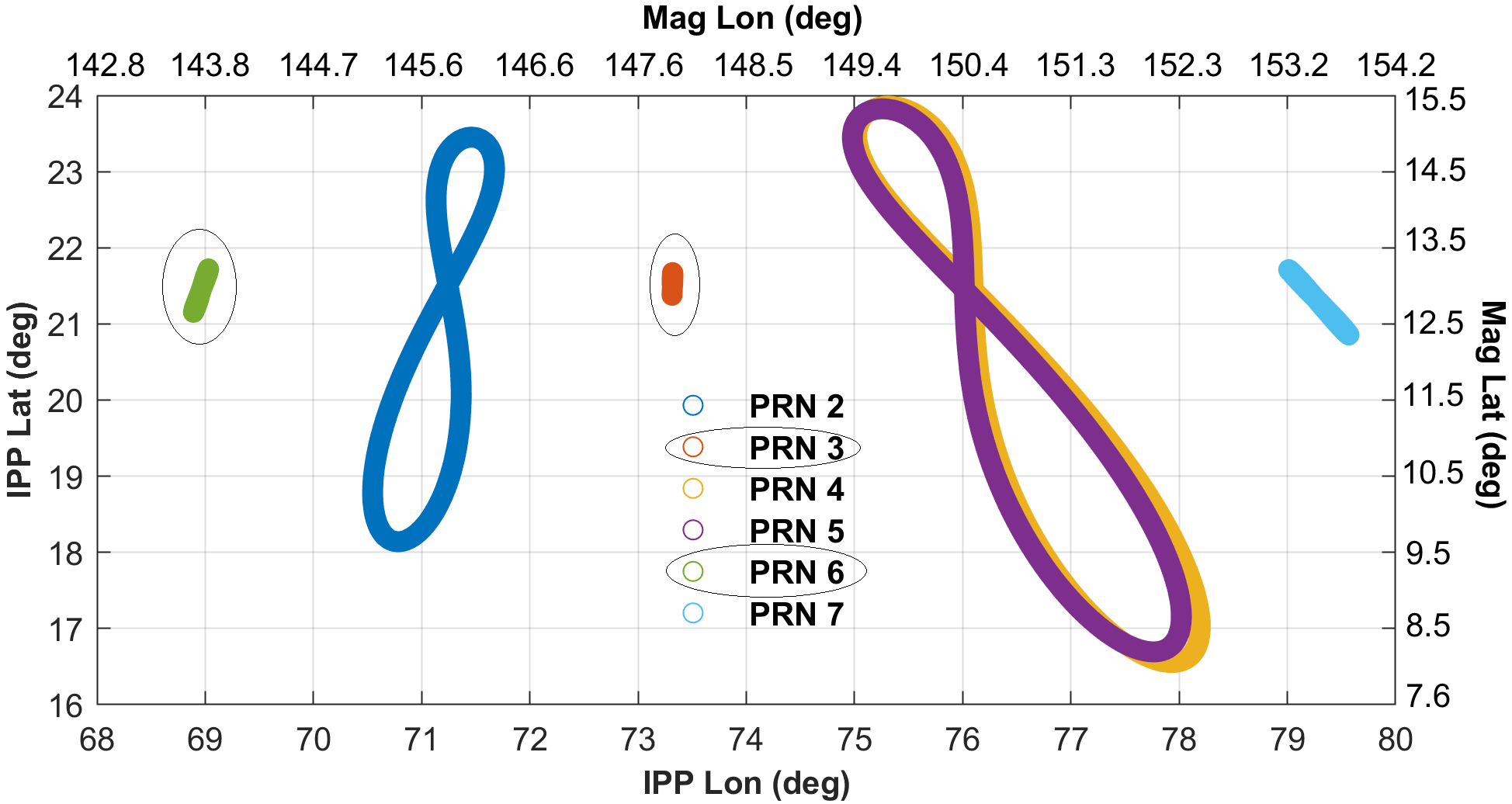}
\caption{Plot showing IPP (geographic and geomagnetic) latitude (Lat) and longitude (Lon) footprints of NavIC satellite PRNs 2-7 over Ahmedabad. PRNs 2, 3, and 6 are GEOs while PRNs 4 5, and 7 are GSOs. The GEO PRNs 3 and 6 used in this work are marked separately in ovals.}
\label{fig1}
\end{figure}

Figure \ref{fig1} shows the Ionospheric Pierce Point (IPP) latitude-longitude (geographic and geomagnetic) plot of the NavIC satellite footprints with respect to the receiver placed at Ahmedabad. Here the two satellites (or PRNs) 3 and 6 of NavIC are considered in the analysis as they are GEOs. PRN 2, also being a GEO, was excluded from the analysis as it contained data gaps during the analysis period. 
It is also to be noted that the IPP separation between PRNs 3 and 6 is about 4.5$^\circ$ pointing to the fact that the ionospheric plasma volume intercepted by PRN 3 is different from that observed by PRN 6.

Next, we obtained the simulation runs using the National Center for Atmospheric Research (NCAR) Thermosphere-Ionosphere-Electrodynamics General Circulation Model (TIEGCM) hosted by the Community Coordinated Modeling Center (CCMC). The TIEGCM is a non-linear, first-principles, and three-dimensional (3D) depiction of the terrestrial ionosphere and thermosphere coupled as a system. The model consists self-consistent solution of the mid-latitude as well as the low-latitude dynamo fields. TIEGCM solves basic physical equations: the continuity, the energy, and the momentum, using a fourth-order, semi-implicit finite difference method on individual pressure surfaces in a vertical grid, for the neutral and the ion species at a typical time-step of 2 minutes. Several studies \citep{sc:30,sc:31,sc:32,sc:33,sc:34,sc:35,sc:29.1,sc:74} show the capability of this model to understand the electrodynamic processes in the ionosphere, thus making it an ideal simulation platform closely reproducing the coupled ionosphere-thermosphere system of Earth. The standard low-resolution grid parameters in the model are the latitudinal extent from -87.5${^\circ}$ to +87.5${^\circ}$ and longitudinal extent from -180${^\circ}$ to +180${^\circ}$; pressure level (altitude-wise) from -7 to +7 in steps of $\frac{H}{2}$ having the lower boundary at around 97 km and upper boundary around 500-700 km (variation as per solar activity condition). The model inputs are F10.7 (daily solar radio flux) along with either the $K_p$ index or the IMFs $B_x$, $B_y$, $B_z$, the solar wind velocity, and solar wind density. The outputs from this model, specified in all three spatial and temporal dimensions, are the height of pressure surfaces (cm); neutral, electron, and ion temperatures (K); the zonal, meridional, and vertical neutral winds (m/s); potential in geomagnetic as well as geographic coordinates; compositions such as O, O${_2}$, NO, N(4S), N(2D), O+, O${_2}$+, N${_2}$+, NO+, N+, Ne.

We have also obtained meridional wind variations from the NCAR Whole Atmosphere Community Climate Model with thermosphere and ionosphere extension (WACCM-X), a self-consistent general circulation model, that fully couples the chemistry and dynamics, and calculates the 3D temperature, ionospheric structures, composition, and wind, from the surface up to 500-700 km in altitude depending on geomagnetic and solar activities \citep{sc:88,sc:89}. The model inputs include solar spectral irradiance, and high-latitude ionospheric inputs, similar to the TIEGCM while the outputs are the ion and electron densities and temperatures, neutral winds and ion drifts (zonal, meridional and vertical), Pederson and Hall conductivities, and compositions (O, O${_2}$, NO, H).   

The high-resolution (1-minute) interplanetary and SYM-H data used in this work are obtained from the OMNI database \\(https://omniweb.gsfc.nasa.gov/) of SPDF, GSFC NASA, the $K_p$ and $A_p$ while the Equatorial Electrojet (EEJ) data are obtained from the Indian Institute of Geomagnetism, Mumbai. The strengths of EEJ are computed by taking the differences of the horizontal (H) component ($\Delta H$) of the Earth’s magnetic field at Alibag (18.6$^\circ$N, 72.9$^\circ$E geographic, dip angle: 26.4$^\circ$, an off-equatorial station), from that at Tirunelveli (8.7$^\circ$N, 77.7$^\circ$E geographic, dip angle: 1.7$^\circ$, a dip equatorial
station). The $\Delta H$ represents daytime instantaneous H values corrected for the nighttime base values of H. The SuperMag Auroral Electrojet (SME) data are obtained from the SuperMag website (https://supermag.jhuapl.edu/) while the global thermospheric O/N$_{2}$ ratio maps are obtained from the Global UltraViolet Imager (GUVI) onboard the Thermosphere-Ionosphere-Mesosphere-Energetics-Dynamics (TIMED) spacecraft hosted by the Johns Hopkins University Applied Physics Laboratory (JHUAPL) website (http://guvitimed.jhuapl.edu/guvi-galleryl3on2). We have also obtained International GNSS Service (IGS) GPS TEC data, available at http://garner.ucsd.edu/pub/rinex/, for the stations Hyderabad (17.41$^\circ$N, 78.55$^\circ$E, geographic; dip angle: 21.69$^\circ$) and Bengaluru (13.02$^\circ$N, 77.50$^\circ$E, geographic; dip angle: 11.78$^\circ$).

\section{Results}

\subsection{Sources of the geomagnetic storms}

According to the Space Weather Prediction Center (SWPC) of the National Oceanic and Atmospheric Administration (NOAA) forecasts (https://www.swpc.noaa.gov/), an X1-class flare, from the Active Region (AR) 2887 at 15:35 UT on October 28, 2021, caused an R3 (strong) level radio blackout. This flare event, having a CME-related signature to it, was the cause of an S1 (minor) radiation storm that began at 17:40 UT on the same day. A minor level geomagnetic storm started around 22:00 UT on October 30, 2021, as a result of this CME interacting with the geomagnetic field.

Following this event that caused a weak geomagnetic storm on October 30-31, 2021, several CMEs erupted from the Sun, with a few of them being Earth-directed, during the period between November 01 and 02, 2021. The source region, of these CMEs, was the southwest area of the Sun which included a C4-class flare from the same AR 2887, erupting at 21:33 UT on November 01, 2021. This event was followed by a full halo CME, from AR 2891 (near the center disk of the Sun) at 03:01 UT on November 02, 2021, related to an M1-class flare that had caused an R1 (minor) level radio blackout. The anticipated Earth-directed CME was observed, by the NOAA Deep Space Climate Observatory (DSCOVR) satellite, as an interplanetary shock at 19:42 UT on November 03, 2021. The CME arrived at Earth in the form of an impulse observed by the GOES-16 satellite, and the ground-based magnetometers at 19:57 UT, thus causing a G3 (strong) level geomagnetic storm by 23:59 UT on the same day. This G3 level storm was preceded by a G1 (minor) level storm at 21:24 UT and a G2 (moderate) level storm at 21:46 UT on November 03, 2021. The geomagnetic storm persisted up to around 13:00 UT on November 04, 2021.

It is important to note that the occurrence of interacting CMEs plays a crucial role in the geoffectivity. In the heliosphere, these interactions evolve in a non-linear manner. In general, if two CMEs get launched in the same direction, the spatiotemporal windows of the different phases of interaction will depend on the ambient solar wind, the mass and initial speed of individual CMEs, magnetic structures and orientations, the time interval between the two individual CME eruptions, the relative speed between the two CMEs \citep{sc:83}. Although most geomagnetic storms are driven by a single CME, about 27$\%$ events \citep{sc:84} are observed to be caused by the passage of interacting CMEs as well as the Stream Interaction Regions (SIRs) which gets formed when the High-Speed Streams (HSS) (about 600-800 km/s and emanating from the coronal holes) catch up with the slower streams (about 300-400 km/s) that are ahead and compress them. Using in situ and remote-sensing observations, an investigation by \citep{sc:85} estimated geoeffectivity amplification (by an almost double factor) of individual CMEs due to CME-CME interactions at 1 AU. Furthermore, interacting CMEs lead to Alfvenic wave fluctuations which in turn may affect the solar wind electric field. Alfvenic oscillations are oscillations of both the motion of the plasma and the magnetic field, with the energy of the wave propagating along the magnetic field lines. Apart from the solar wind and the solar atmosphere, these oscillations are also found in the magnetosphere of our planet, arising from the disturbances within the magnetosphere itself or from the solar wind-magnetosphere interaction. \cite{sc:86} suggested that the interaction or collision between the CME and HSS deforms the MC of the CME which in turn generates Alfven waves inside the MC.

\subsection{Interplanetary and Geomagnetic conditions}

\begin{figure}[ht]
\centering
\includegraphics[scale=0.44]{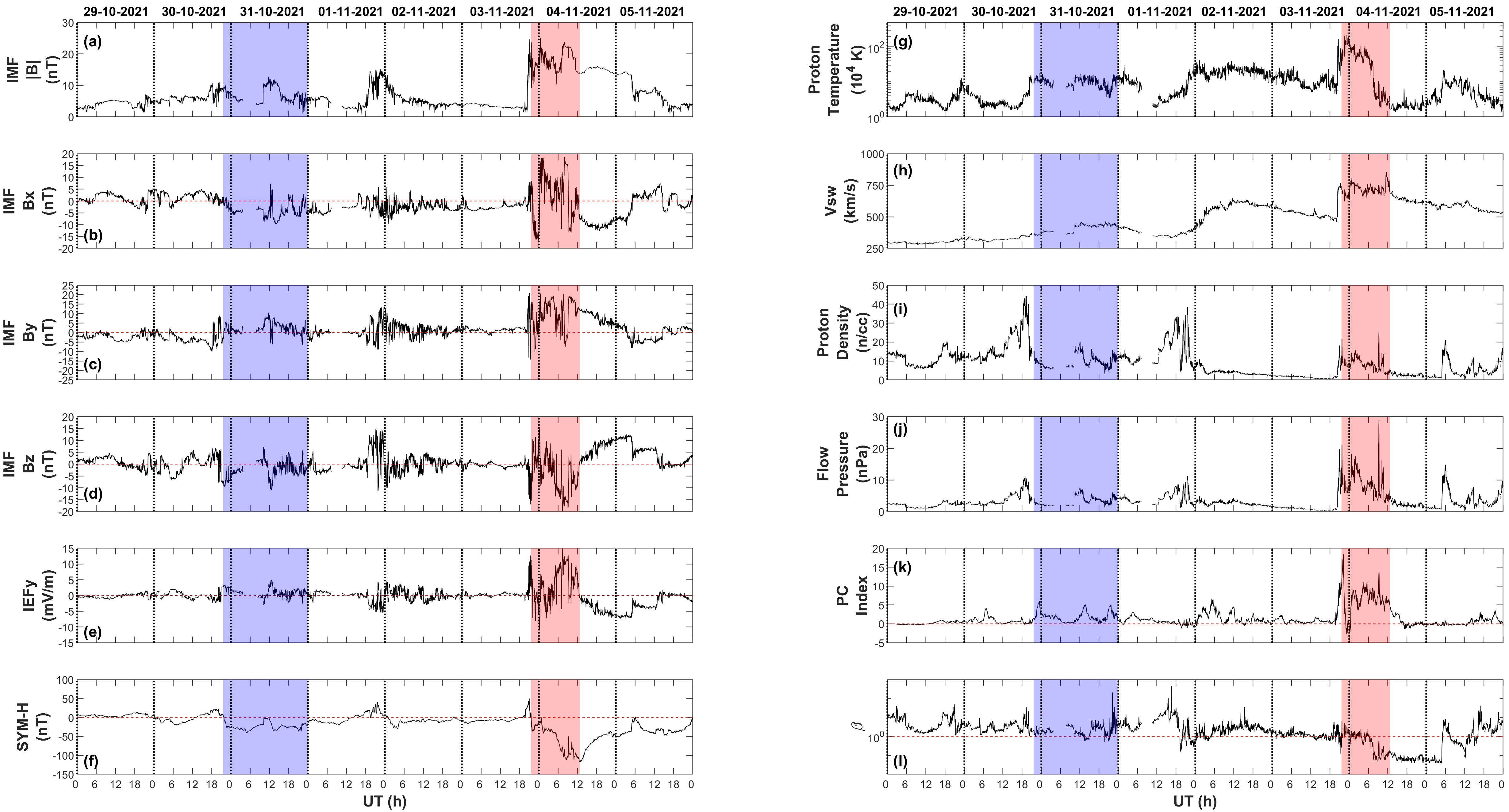}
\caption{Figure panels (a) to (f) show the IMFs $|B|$, $B_x$, $B_y$, $B_z$, the $IEF_y$ and the SYM-H (nT) variations respectively. In the same figure, panels (g) to (l) show the Proton Temperature, the $V_{sw}$, the Proton Density, the Flow Pressure, the PC index, and the $\beta$ respectively between October 29 and November 05, 2021. The period of the weak and the strong geomagnetic storm occurrences are shown as violet- and peach-shaded regions respectively.}
\label{fig2}
\end{figure}

Figure \ref{fig2} shows the SYM-H and the interplanetary conditions between October 29 and November 05, 2021. Figure \ref{fig2}: panels (a) to (d) show the IMFs $|B|$, $B_x$, $B_y$, $B_z$ (nT) variations respectively, panels: (e) and (f) show the Interplanetary Electric Field ($IEF_y$, mV/m) and the SYM-H (nT) variations respectively. In the same figure, panels: (g) to (l) respectively show the Proton Temperature (10$^{4}$ K), the Solar Wind Flow Speed ($V_{sw}$, km/s), the Proton Density (n/cc), the Flow Pressure (nPa), the Polar Cap (PC) index and the Plasma Beta ($\beta$). The periods of weak geomagnetic storm occurrences are shown as the violet-shaded region while the sheath region and the corresponding strong geomagnetic storm occurrence are shown as the peach-shaded region in this figure. The coverages (LT = UT + 5h) of the main phase of the weak geomagnetic storms are from 02:36 to 10:00 LT on October 31, 2021, and from 16:52 on October 31 to 01:59 LT on November 01, 2021, while the same for the stronger event is from 02:36 to 17:45 LT on November 04, 2021.

The Storm Sudden Commencement (SSC) of the weak storm occurred at 20:05 UT (01:05 LT) on October 30 (31), 2021. The main phase of this relatively weaker (compared to the storm occurring on November 03-04, 2021) storm started at 21:36 UT (02:36 LT) on October 30 (31), 2021. The corresponding values of IMFs $|B|$, $B_x$, $B_y$, $B_z$, $IEF_y$, Proton Temperature, $V_{sw}$, Proton Density, Flow Pressure, PC index and $\beta$ were 8.04 nT, 1.11 nT, -2.98 nT, -7.37 nT, 2.72 mV/m, 9.6317*10$^{4}$ K, 368.90 km/s, 14.64 n/cc, 3.98 nPa, 1.29 and 2.12 respectively. The SYM-H dropped to a minimum with the value of -41 nT at 05:00 UT (10:00 LT) on October 31, 2021. Next, a second weak storm occurred when the main phase started at 11:52 UT (16:52 LT) on October 31, 2021, with the corresponding IMFs $|B|$, $B_x$, $B_y$, $B_z$, $IEF_y$, Proton Temperature, $V_{sw}$, Proton Density, Flow Pressure, PC index and $\beta$ values of 10.54 nT, -4.68 nT, 5.76 nT, -7.46 nT, 3.24 mV/m, 13.8349*10$^{4}$ K, 433.90 km/s, 19.85 n/cc, 7.47 nPa, 1.63 and 1.98 respectively. It is to be noted that during this period, the SYM-H had two minimum peaks with values of -36 nT and -37 nT at 13:24 UT (18:24 LT on October 31, 2021) and 20:59 UT (01:59 LT on November 01, 2021) respectively. The corresponding values of IMFs $|B|$, $B_x$, $B_y$, $B_z$, $IEF_y$, Proton Temperature, $V_{sw}$, Proton Density, Flow Pressure, PC index, and $\beta$ during these two SYM-H drops were (10.67 nT, -7.65 nT, 7.22 nT, -1.73 nT, 0.75 mV/m, 12.2953*10$^{4}$ K, 432.40 km/s, 10.31 n/cc, 3.86 nPa, 4.69 and 0.95 respectively) and (7.93 nT, -6.39 nT, -0.49 nT, -4.67 nT, 2.11 mV/m, 4.9839*10$^{4}$ K, 452.70 km/s, 6.16 n/cc, 2.52 nPa, 3.37 and 0.73 respectively). 

Further on observing the peach-shaded region depicting the strong geomagnetic storm, the SSC had occurred at 19:44 UT (00:44 LT) on November 03 (04), 2021. The main phase of the strong storm started at 21:36 UT (02:36 LT) on November 03 (04), 2021, while the SYM-H reached its minimum value of -118 nT at 12:45 UT (17:45 LT) on November 04, 2021. The corresponding values of IMFs $|B|$, $B_x$, $B_y$, $B_z$, $IEF_y$, Proton Temperature, $V_{sw}$, Proton Density, Flow Pressure, PC index, and $\beta$ during the main phase onset and the SYM-H minimum were (23.33 nT, 6.79 nT, 15.81 nT, -15.68 nT, 11.23 mV/m, 41.0965*10$^{4}$ K, 716.10 km/s, 7.66 n/cc, 7.86 nPa, 10.34 and 0.32 respectively) and (14.00 nT, -7.60 nT, 11.73 nT, 0.71 nT, -0.49 mV/m, 1.7323*10$^{4}$ K, 686.10 km/s, 4.83 n/cc, 4.55 nPa, 4.32 and 0.15 respectively).  

Comparing the two shaded regions, one can observe enhanced IMF, Proton Temperature, Density, and Pressure variations in addition to enhanced values of solar wind velocities, for the peach-shaded stronger event (November 04, 2021). These observations suggest a highly compressed and turbulent plasma and clearly show the presence of a sheath region during this period. Since these sheath regions are highly turbulent as a result of plasma instabilities, they are capable of triggering substorms, which in turn affects the auroral and high-latitude as well as low/equatorial ionosphere (see \cite{sc:80,sc:81,sc:71,sc:82} and references therein).

\subsection{Ionospheric conditions}

The ionospheric TEC is known to vary on a daily basis, with time, over different seasons and locations (latitude and longitude) as well as with varying solar activity levels. Over the Indian longitude sector and under quiet-time conditions, the daytime values of TEC are observed to be higher during equinoctial months (Vernal: March and April; Autumnal: September and October) followed by the winter (November to February) and the summer (May to August) months. These seasonal trends are controlled by the thermospheric neutral wind that changes the neutral composition as well as increases the O/N${_2}$ ratio over the low/equatorial latitudes, attaining peak values over the equinoctial periods. This increase in the O/N${_2}$ ratio causes increases in the electron density and corresponding higher TEC values are observed during the equinoctial months \citep{sc:91}. However, under disturbed conditions, significant variations from these general trends can be observed. For example, in the recent study by \citep{sc:92}, they highlighted that quiet-time seasonal trends are not observed during disturbed periods, especially under moderate solar activity conditions, a higher difference in TEC (from the quiet-time values) can be observed and there are also reductions in the correlation between the solar flux and the diurnal maximum TEC under storm-time conditions.

Under quiet-time conditions, it is well known that increases in the upward plasma motion, and hence the plasma getting raised to higher altitudes, are caused by enhancements in the E $\times$ B vertical drifts. In this process, redistribution of plasma occurs from the equatorial regions to the regions in and around the EIA crest. However, under storm-time conditions, the PPEF produces changes in the low-latitude ionosphere by lifting the equatorial ionosphere to higher altitudes while decreasing the plasma density at the bottomside ionosphere. These F-region uplifts that are caused by the PPEFs, further increase the altitudes of these plasma. This process reduces the rate of recombination which in turn leads to high electron densities and intensification in the TEC over the EIA crest regions \citep{sc:87}. In the current study, due to the presence of the MC region and subsequent stable IMF $B_z$, we observe higher values of TEC during a weaker event as compared to the stronger one when the IMF $B_z$ had been strongly fluctuating due to the presence of the sheath region. The observations of such anomalous response of the low-latitude ionosphere (over quiet-time values) to MC-like and sheath-region-induced disturbances, are described next.

\begin{figure}[ht]
\centering
\includegraphics[scale=0.65]{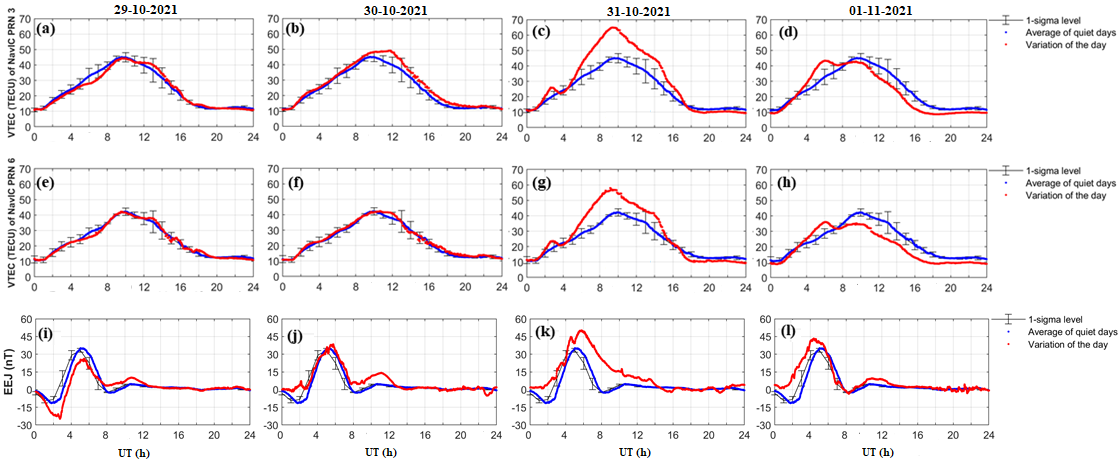}
\caption{Variations of NavIC VTEC for PRN 3 (panels (a), (b), (c), (d)), for PRN 6 (panels (e), (f), (g), (h)), along with the EEJ variations (panels (i), (j), (k), (l))} during the period from October 29 to November 01, 2021. The red curves show the observed VTEC and EEJ variations while the blue curves show the variation of the average of the three quietest days. The black horizontal bars signify the 1-sigma level.
\label{fig3}
\end{figure}

\begin{figure}[ht]
\centering
\includegraphics[scale=0.65]{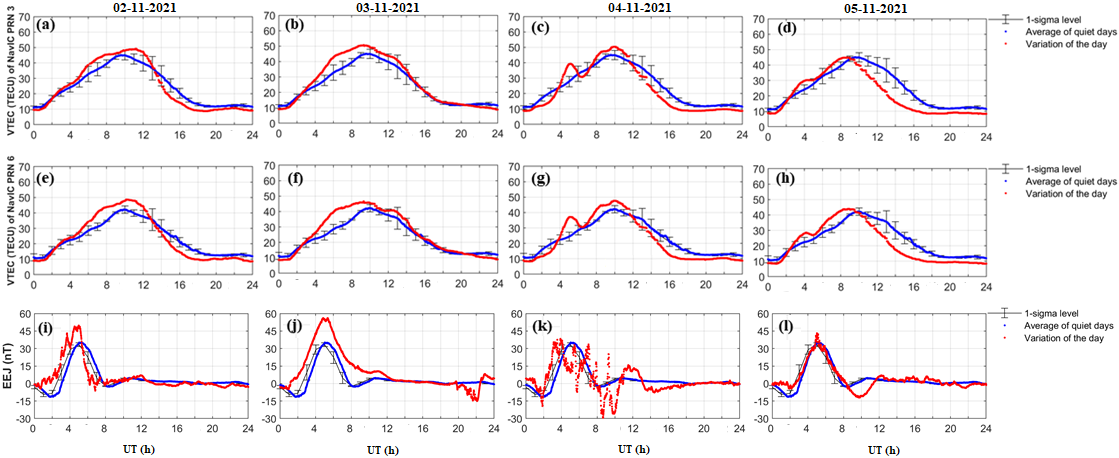}
\caption{Same depiction as Figure \ref{fig3} but observed during November 02-05, 2021.}
\label{fig4}
\end{figure}

Figures \ref{fig3} and \ref{fig4} show the ionospheric diurnal VTEC variations as observed by the NavIC PRNs 3 and 6 along with the EEJ variation during the entire analysis period. The titles on top of the first panels in both the figures show the corresponding days in the format: DD-MM-YYYY. In all the panels of these two figures, to observe significant differences/enhancements from the day-to-day variability, the average of the three quietest days' VTEC and EEJ values are plotted as blue curves. The three quietest days selected here are October 27, 28, and 29, 2021, based on the availability of both the TEC and the EEJ data around the analysis period. The observed VTEC variations on the individual days are plotted in red curves. The horizontal bars in these figures show the 1-sigma level to designate the VTEC and EEJ variations, beyond these limits, to be significant due to solar forcing and not due to the daily variabilities. 

Figure \ref{fig3} panels (a) to (l) show these variations from October 29- November 01, 2021. The day of October 29, being one of the quietest days shows the VTEC variations to be well within the significance bars and close to the quiet-time variation as observed by PRNs 3 (panel (a)) and 6 (panel (e)). Coming to panels (c) and (g) of the same figure, it can be noted that there have been significant rises in the VTEC values as observed by the NavIC PRNs, on October 31, 2021. This is the same day when there had been successive weak geomagnetic storm occurrences (see Figure \ref{fig2} panel (d)). The diurnal maximum around 09:30 UT (14:30 LT) showed enhancements of about 20 TECU over the quiet-time variation, as observed by PRN 3. Interestingly, PRN 6 showed a significant enhancement but with a lower value of about 15 TECU over the quiet-time values around the same time. This could be attributed to the fact that the IPP separation between these PRNs is about 4.5$^\circ$ and hence the ionospheric plasma volumes intercepted by these PRNs would be different. For the day (November 01, 2021) after the event day (October 31, 2021), panels (d) and (h) for PRNs 3 and 6 respectively, show no significant enhancements in the VTEC. Coming to Figure \ref{fig4}, no significant enhancements outside the 1-sigma limits in the VTEC variations are observed in panels (c) and (g) for PRN 3 and PRN 6 on November 04, 2021, respectively, despite the occurrence of the strong geomagnetic storm (see panel (d) of Figure \ref{fig2}). Additionally, the observation of a lesser geoeffectivity on November 04, can be attributed to the presence of Alfvenic wave fluctuations as a result of interacting CMEs (with the second CME having a higher velocity than the first and assuming no acceleration and/or deceleration in the interplanetary medium) that occurred during November 01-02, 2021.

\vspace{20pt}
Next, Panels (i), (j), (k), and (l) in Figures \ref{fig3} and \ref{fig4} show the corresponding EEJ variations from October 29- November 01, 2021, and November 02-05, 2021 respectively. On October 31, 2021, there was an observation of enhancement in the EEJ values over the quiet-time variation (panel (o) in Figure \ref{fig3}). However, on observing the same on November 04, 2021 (panel (o) of Figure \ref{fig4}), the occurrences of counter-electrojets (CEJ) can be observed which might have restricted the development of the plasma fountain over the crest region on that day, thereby, showing no significant TEC enhancement in comparison to the weaker events. This CEJ was caused as a result of sudden reversals in the IMF $B_z$ (see panel (a) of Figure \ref{fig2}) orientation from southward-to-northward during certain intervals of time as a result of the presence of ICME-sheath region. Also from Figure \ref{fig2}, we can observe the main phase of the weak storm on October 31, which started at 02:36 LT and ended at 10:00 LT. This is also the time when we observe the strengthening of the EEJ over this sector, which in turn strengthens the TEC over the crest regions at around 14:30 LT. In the case of the strong event on November 04, the main phase occurred from around 02:30 LT to 17:45 LT, covering the entire daytime of November 04. However, we can observe strong CEJ and hence not so-prominent strengthening of the plasma fountain over the crest region and subsequent lower values of TEC on that day. Therefore, it becomes important to note that although the magnitude of the southward IMF had caused a strong geomagnetic storm, evident from the intensification of the ring current (lower SYM-H values), it failed to be significantly geoeffective due to the absence of a long-duration southward IMF $B_z$ during this period. 

\begin{figure}[ht]
\centering
\includegraphics[scale=0.65]{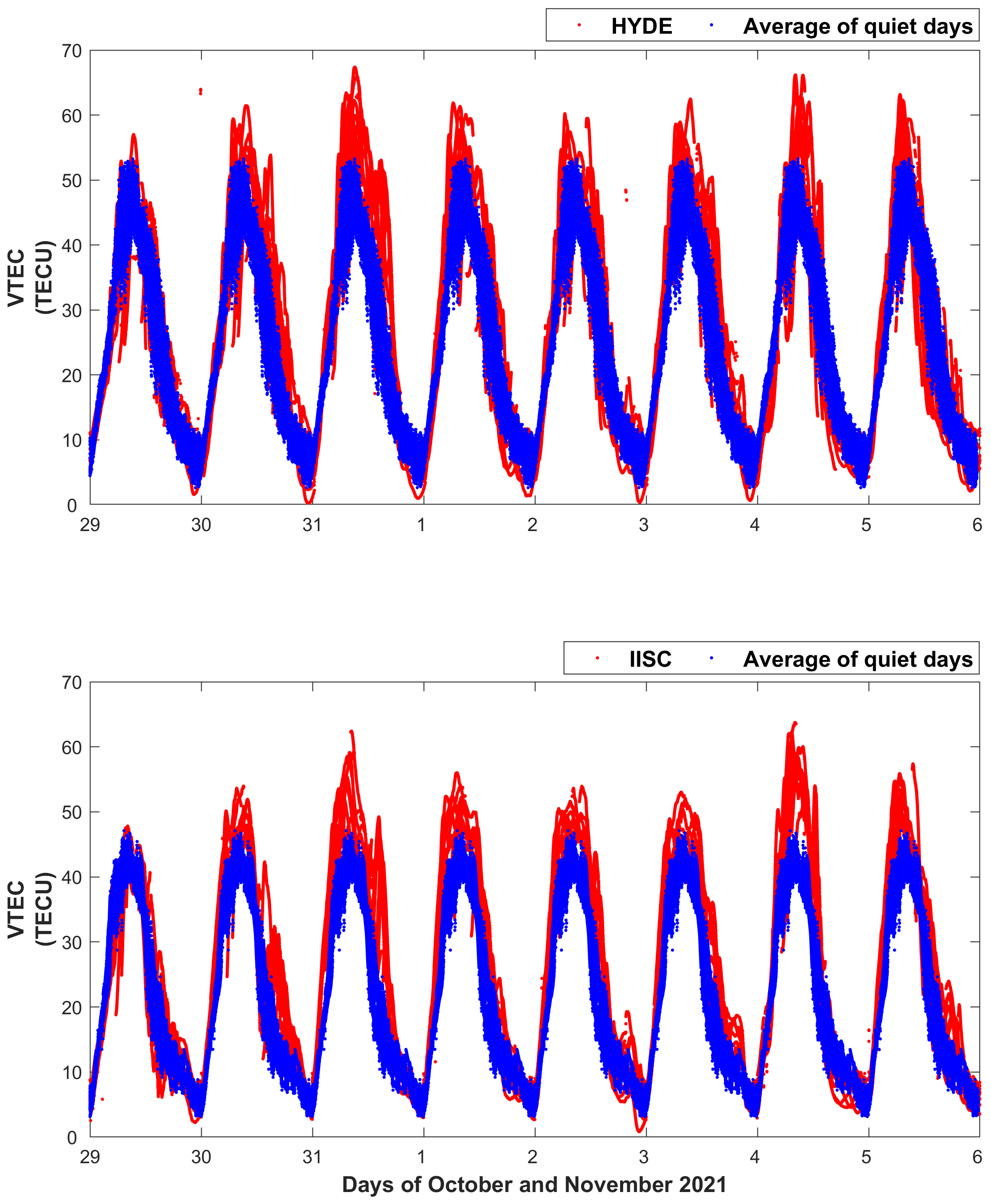}
\caption{IGS GPS TEC variation over Hyderabad (HYDE, Top) and Bengaluru (IISC, Bottom) observed from October 29 to November 05, 2021. The red curves show the observed VTEC while the blue curves show the variation of the average of the three quietest days.}
\label{fig5}
\end{figure}

Next, Figure \ref{fig5} shows the GPS TEC variations over the stations Hyderabad (HYDE, Top Panel) and Bengaluru (IISC, Bottom Panel) for the entire analysis period between October 29 and November 05, 2021. The red curves show the VTEC variations on the individual days while the blue curves show the variations of the average of the three quietest days. Looking into the top panel, for HYDE, we can observe the diurnal maximum of TEC (red curve) on October 31 (third bell curve) to be about 68 TECU which is about 16 TECU above the quiet-time variation (blue curve). The variation on November 04 (second last bell curve), showed the same at around 65 TECU. Looking into the bottom panel, for IISC, we see that the diurnal maximum on October 31 and November 04 is around 62-63 TECU. From these panels, we can say that there is an increase in the diurnal maximum of TEC (from 62 TECU at IISC to about 68 TECU at HYDE) on October 31, as one approaches the EIA crest (Bengaluru to Hyderabad). This can be attributed to the presence of the effects of the neutral winds that could have played a role, in tandem with the non-fluctuating PPEF, in causing a higher variation in TEC over this location and Ahmedabad on October 31, 2021. The following paragraph discusses this aspect in detail. 

\begin{figure}[ht]
\centering
\includegraphics[scale=0.75]{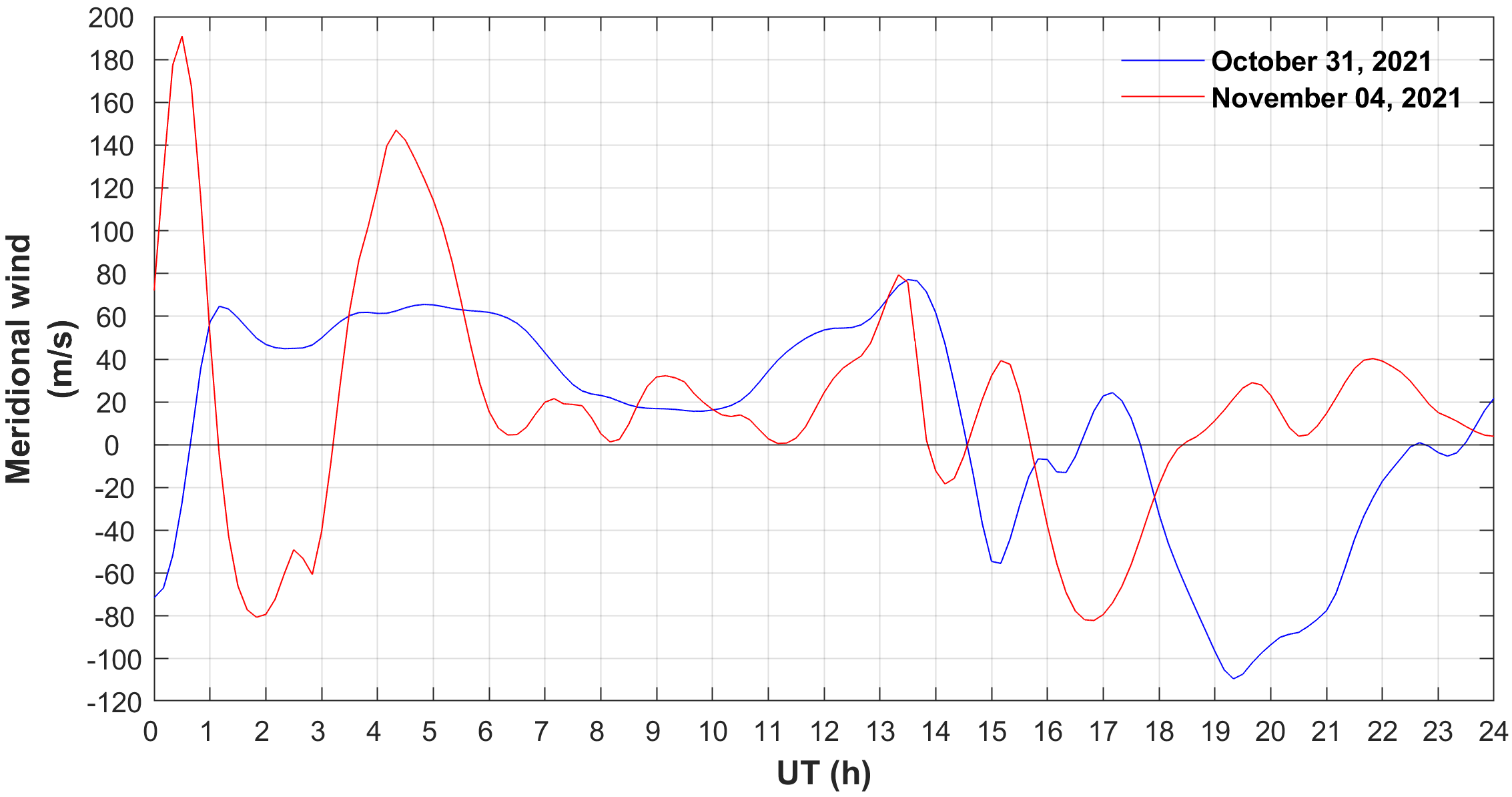}
\caption{Meridional wind (m/s) variations (positive signifying poleward-directed wind) from WACCM-X during October 31 (blue) and November 04 (red), 2021.}
\label{fig6}
\end{figure}

Now, the neutral dynamics such as the thermospheric winds play a role in the modulation of the responses of the ionosphere during storm-time conditions. These winds contribute to the equatorial E $\times$ B plasma drift causing variations in the ion-recombination rate and alterations of the TEC \citep{sc:41,sc:42}. In general, the poleward winds move the plasma down to the lower altitudes where the recombination rate is comparatively large. This in turn results in the reduction in the peak height (hmF2) and hence a decrease in the peak electron density \citep{sc:46}. Studies by (see \cite{sc:25} and references therein) have shown that the disturbances in the ionosphere are a combination of several processes or mechanisms, where the dominating mechanism and its significance might vary from one event to the other. Therefore, to show any possible effects of the neutral dynamics on TEC, in addition to the effects of the PPEF, during October 31 and November 04, 2021, the variations from the WACCM-X simulated 350 km meridional wind velocities (m/s) over the observation site is shown in Figure \ref{fig6}. In this Figure, one can observe a higher magnitude of the poleward wind on November 04 compared to October 31 between 9 and 10 UT (which is the period around the observed diurnal maximum of TEC). Specifically, on October 31, at 9:30 UT, wind velocities are about 16 m/s whereas at the same time, on November 04, 2021, wind velocities are close to 32 m/s. We believe that this difference might have played some role, in tandem with the PPEF, that had caused the anomalous variations in TEC on October 31 and November 04, 2021.

While using the wind estimates from the WACCM-X model, one may wonder how well correlated the model outputs are compared to actual observations. In a study by \citep{sc:93}, they compare WACCM-X simulations with the WINDII observations of high-latitude wind under different solar radio flux levels and varying geomagnetic conditions. They conclude that the WINDII observations are in good agreement with the WACCM-X estimates. However, the magnitude of the wind estimate from the model is lower than that of observations, although the wind variation is quite similar. This underestimation of the model could be attributed to the fact that WACCM-X is a climatological model and the abrupt changes in the wind velocity under different geomagnetic disturbances, may not always be captured. However, since WACCM-X is a self-consistent general circulation model that couples the chemistry and the dynamics from the surface to about 700 km altitude, and is capable of reproducing the ionosphere-thermosphere state, as well as the variabilities, of the whole atmosphere, on an hourly to daily time scales due to both geomagnetic and lower atmosphere forcing, the general trends are expected to be similar to observations \citep{sc:88,sc:89}.

\begin{figure}[ht]
\centering
\includegraphics[scale=1]{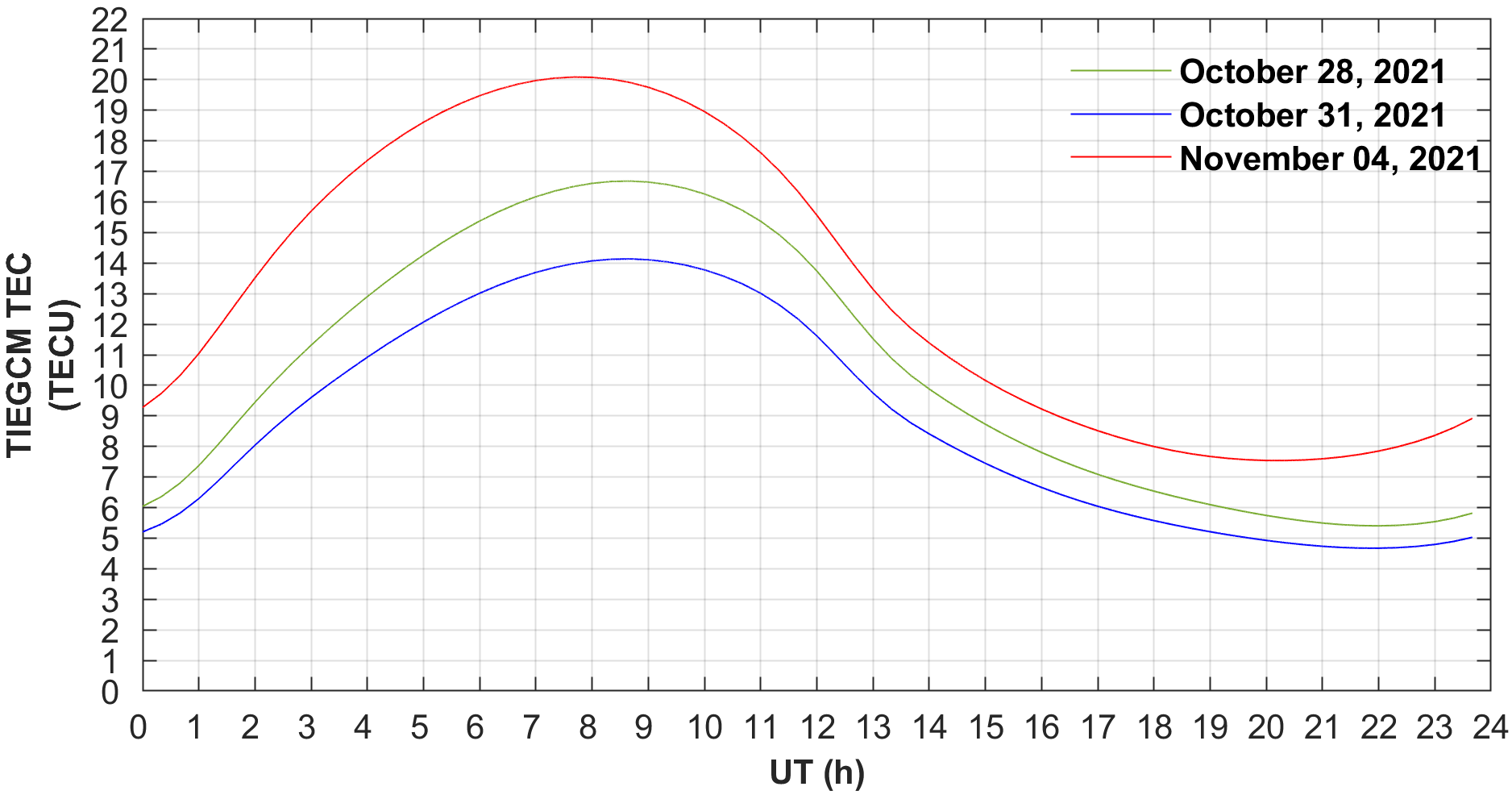}
\caption{TIEGCM simulation results showing the TEC variations over Ahmedabad on a quiet day (October 28, 2021: green curve), on October 31, 2021 (blue curve), and on November 04, 2021 (red curve).}
\label{fig7}
\end{figure}

Next, Figure \ref{fig7} has been shown to check whether the observed anomalous behavior is captured by a global circulation model. The TIEGCM is used to obtain TEC variations on a quiet day (October 28, 2021: green curve), the day (October 31, 2021: blue curve) when successive weak storms had occurred, and the day (November 04, 2021: red curve) when a strong geomagnetic storm had occurred. The inputs given to the model are the F10.7 and the $K_p$ index. The TIEGCM is known to produce consistent results in the coupled ionosphere-thermosphere system, however, in the present case, it produces just the opposite behavior. Table \ref{tab1} shows the observed diurnal maximum TEC as well as the model output on each of these three days. When we compare the variation of TEC on October 31 and November 04 with October 28 (quiet day), we note that lower TEC is observed on October 31 and higher TEC is observed on 04 November. The overall observations show that the TIEGCM is unable to capture the enhanced response of the low-latitude ionosphere on October 31, 2021. Despite TIEGCM's consistency in reproducing ionospheric-thermospheric conditions, the discrepancies observed between the model outputs and the observed data for these three cases could be attributed as follows. The TIEGCM has dependencies on the solar radio flux (F10.7, s.f.u) with either the $K_p$ index or the combination of the three IMF components, the solar wind velocity, and density. For the case of October 28, the F10.7 was comparatively higher (110.3) than that on October 31 (101.2), while the other parameters were comparable. As a result, we observed a slightly higher variation of TEC on October 28 than on October 31, 2021. Further, the model is expected to show the highest variations on November 04, 2021, as the solar wind parameters as well as the $K_p$ values are higher compared to the other two days. Hence we see an overall rise in the TEC variation on that day. As our observations show an anomalous (strong ionospheric responses under weak conditions) case, therefore, incorporation of such types of anomalous cases and the corresponding ionospheric responses, in the model, could be beneficial and valuable to the space sciences community in terms of the improvements of the model's performance and enhancing its predictive capabilities.

\begin{table*}
\centering
\vspace{20pt}
\caption{Diurnal maximum TEC variations from observations and model outputs for October 28, October 31, and November 04, 2021.}
\vspace{10pt}
\begin{tabular}{|l|l|l|l|}
\hline
Days of 2021 &  PRN 3 & PRN 6 & TIEGCM \\
\hline
October 28   &  43.73 &  42.55 &  16.68 \\
\hline
October 31   &  64.51 &  58.25 &  14.14 \\
\hline
November 04  &  50.75 &  46.42 &  20.09 \\
\hline
\end{tabular}
\label{tab1}
\end{table*}

\begin{figure}[ht]
\centering
\includegraphics[scale=0.75]{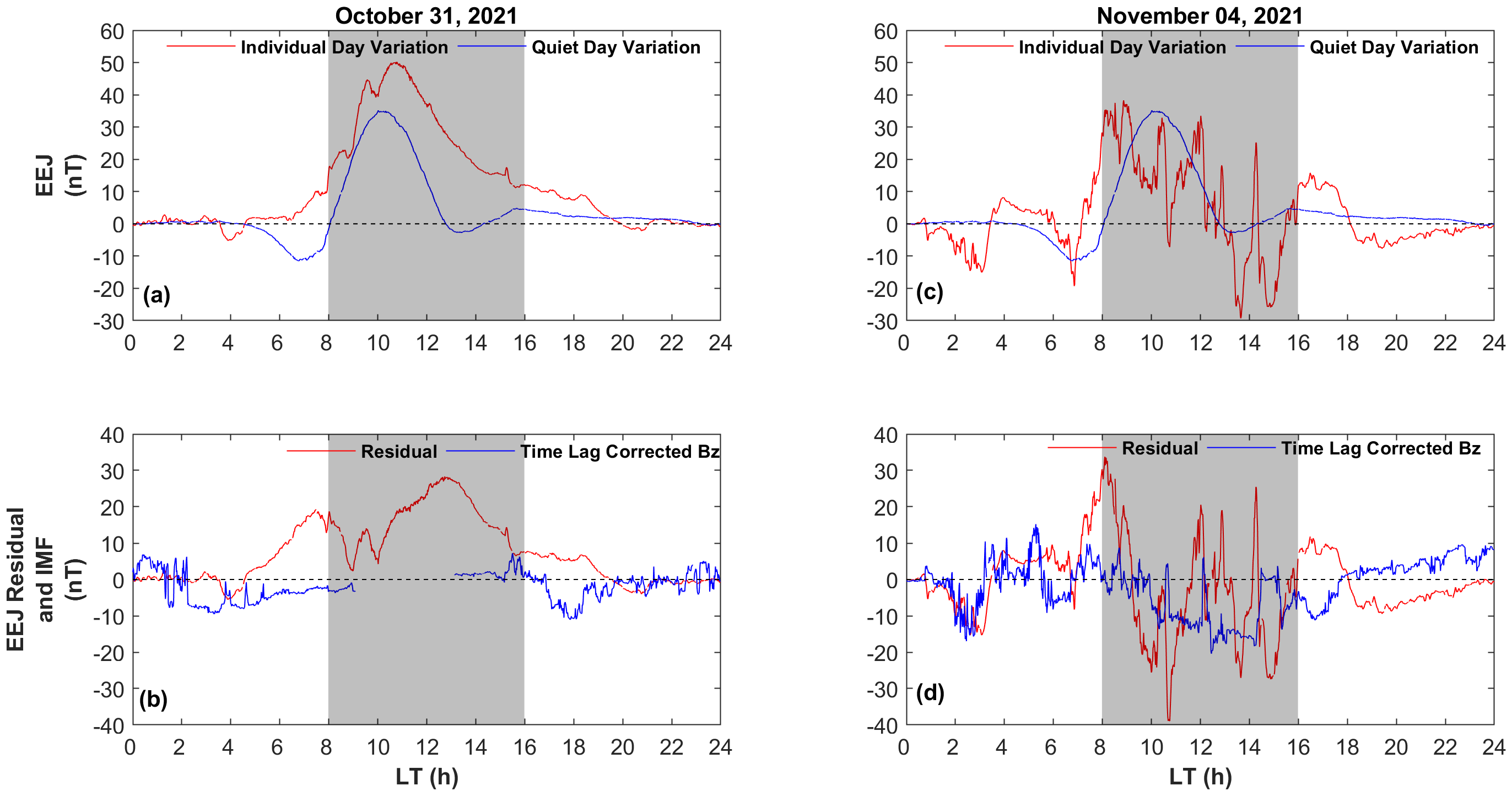}
\caption{Individual day (red curve) along with quiet time variation (blue curve) of EEJ on (a) October 31, 2021, and (c) November 04, 2021. Variations of} residual EEJ (red curve) and time lag corrected IMF $B_z$ (blue curve) on (b) October 31, 2021, and (d) November 04, 2021. The grey-shaded regions show the duration from 08:00 to 16:00 LT for the days.
\label{fig8}
\end{figure}

Figure \ref{fig8} has been plotted to understand the effects of the ICME-driven MC and the sheath structures on the low latitude ionosphere on October 31 and November 04, 2021. In this figure, panels (a) and (c) are the same as the ones plotted in Figures \ref{fig3} and \ref{fig4}, the only difference here being variation has been shown with respect to LT. The shaded region in grey designates the LT hours from 08:00 to 16:00. Panels (b) and (d) show the variations of the residual EEJ and the time lag corrected IMF $B_z$. The residual EEJ is a simple subtraction of quiet-time variation (blue) from the individual day variation (red). The IMF $B_z$ variation, which is shifted to the bow-shock nose, from the spacecraft at the Lagrangian point-L1 has been further shifted to the ionosphere for a one-to-one mapping of the same with the residual EEJ variation. The procedure followed in this work for obtaining the time lag corrected IMF $B_z$ has been thoroughly explained in (see \cite{sc:72} and references therein). Looking into panel (b), one can observe a fluctuating (or AC) penetration electric field, thereby causing a net positive effect on the EEJ and a corresponding low-latitude ionospheric response in terms of TEC enhancements. On the other hand, panel (d) clearly shows the effects of AC penetration fields and the subsequent inhibition effect in the development of the EEJ and negligible effects on the low-latitude TEC variations. 

\begin{figure}[ht]
\centering
\includegraphics[scale=0.75]{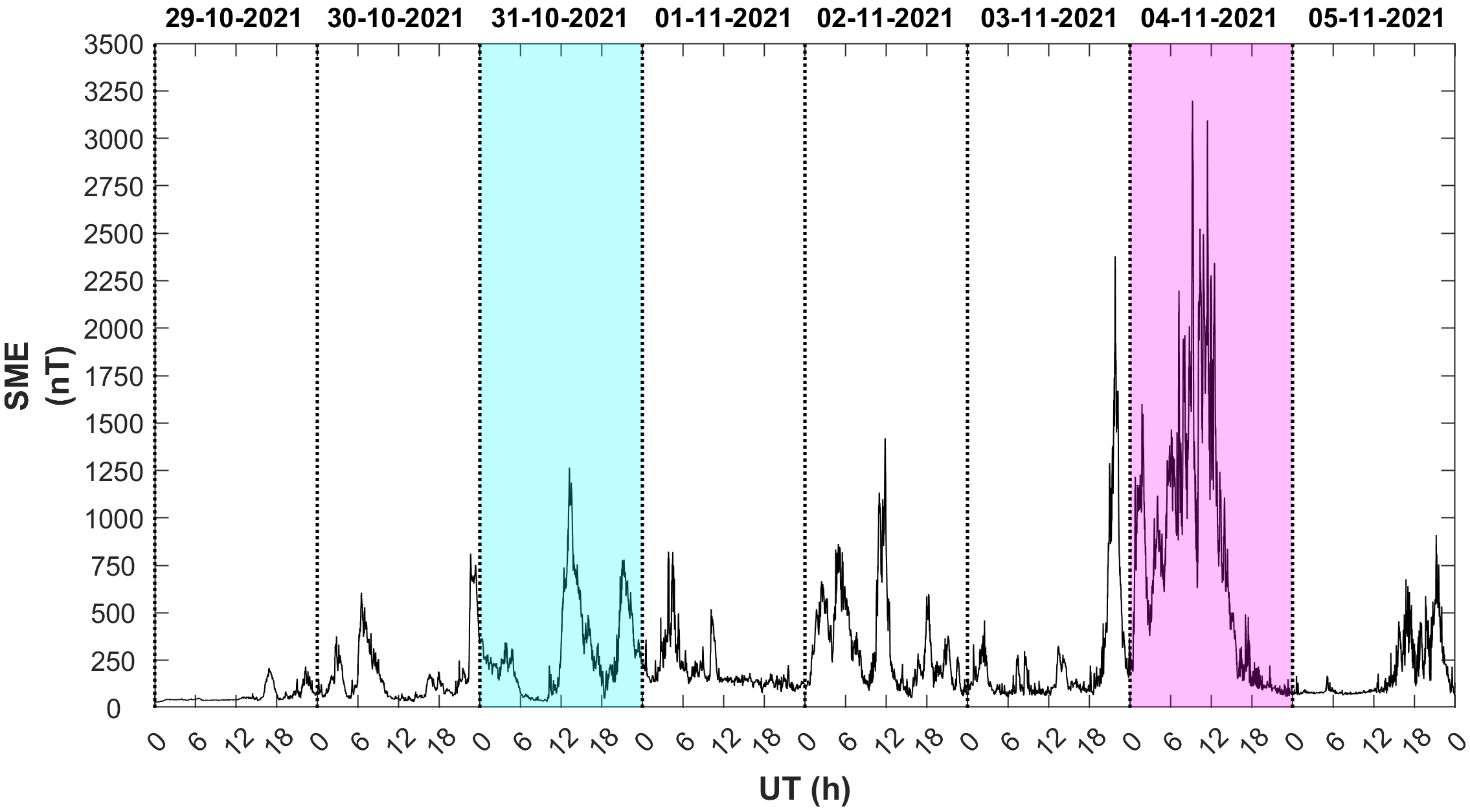}
\caption{The SME (nT) variation from October 29 to November 05, 2021. The cyan-shaded region indicates the day of October 31, 2021, while the pink-shaded region indicates the day of November 04, 2021.}
\label{fig9}
\end{figure}

\begin{figure}[ht]
\centering
\includegraphics[scale=0.45]{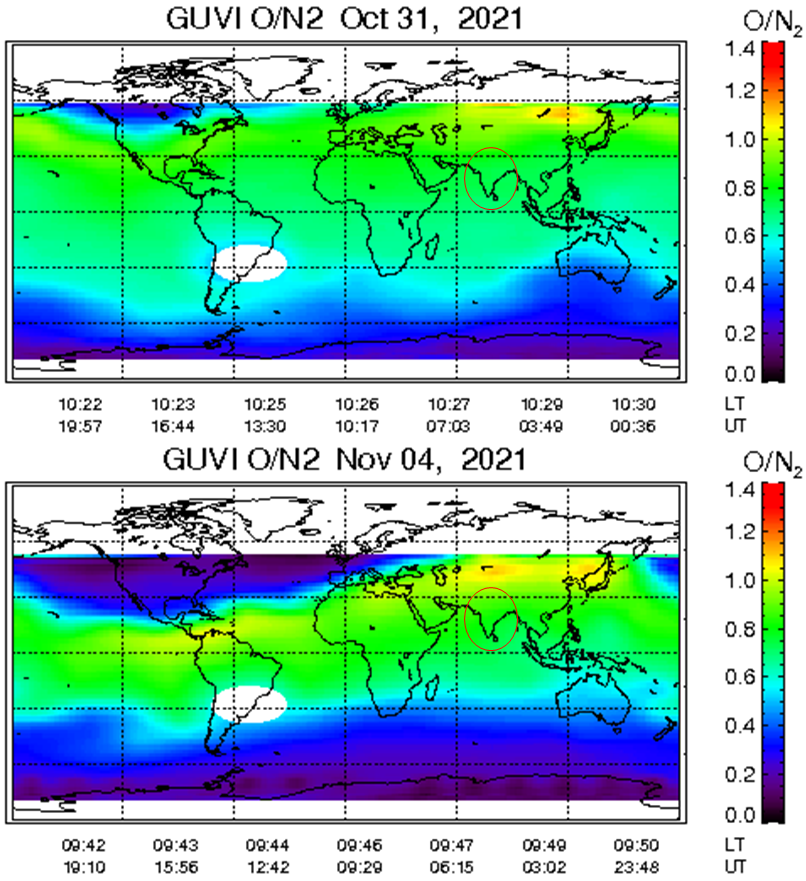}
\caption{The GUVI O/N${_2}$ global maps on October 31 (top), and November 04 (bottom), 2021. The red circles marked on these maps show the Indian region.}
\label{fig10}
\end{figure}

Next, to rule out effects on the observations due to the substorm-induced electric fields and the compositional changes, Figures \ref{fig9} and \ref{fig10} are shown. In Figure \ref{fig9}, we show the variations of the SME (nT) index, for the entire analysis period. The value of SME can be observed to show a rise up to 750 nT around 23:30 UT on October 30, 2021, and a second rise, higher in magnitude (about 1200 nT) around 13:30 UT on November 31, 2021, can be observed in the figure. The SME values start rising again and peak with a value around 3250 nT on November 04, 2021. This suggests that the substorm-induced electric field did not affect the low-latitude ionosphere on the weaker event day significantly. 

Furthermore, in Figure \ref{fig10}, we show the thermospheric O/N${_2}$ ratio maps observed by the GUVI onboard TIMED spacecraft on October 31, and November 04, 2021. An overall value of around 0.75 throughout the Indian region, suggests no significant changes in the O/N${_2}$ values over these regions on both event days. Further explanations of these results will be discussed later.

\clearpage
\section{Discussion}

Before proceeding with the discussion of the physical mechanism that had caused the observed anomalous enhancement of the low-latitude ionosphere during the weaker events, it is important to rule out the effects of substorm-induced electric fields and compositional changes on the low-latitude electrodynamics and ionization. 

In Figure \ref{fig4}, one can observe the occurrence of a CEJ that restricts the development of the fountain on November 04, 2021, and hence no significant enhancements in the ionization on that day. This could be attributed to the fact that the substorm-induced electric field (\cite{sc:47} and references therein) associated with the SME (Figure \ref{fig9}) showing much higher values of around 3250 nT, meant the possibility of much stronger substorm effects on that day. Substorms have the capability of creating reversed electric fields, thereby causing an overshielding condition over the low latitudes. This in turn prevents the increase or strengthening of the EEJ and the corresponding ionization over the northern crest of EIA. On the contrary, during the weaker event, we observed a comparatively lower SME (almost one-third of that observed on November 04, 2021) value which would not be capable of perturbing the low-latitude ionospheric electrodynamics during October 31, 2021.

\begin{figure*}[ht]
\centering
\includegraphics[scale=0.75]{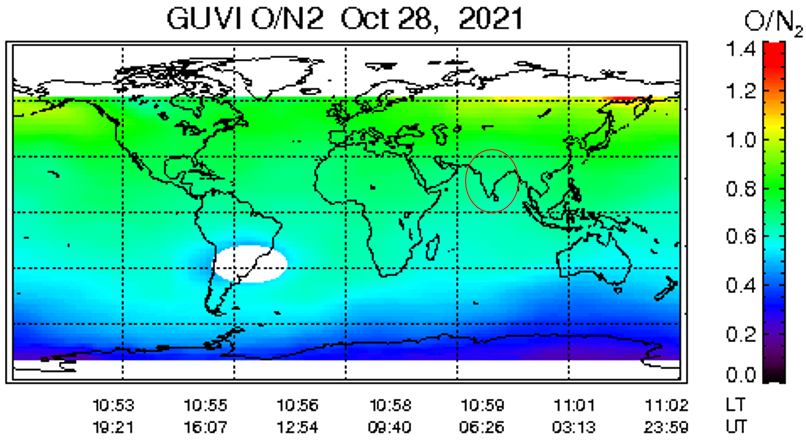}
\caption{Same depiction as Figure \ref{fig9} but observed on October 28, 2021.}
\label{fig11}
\end{figure*}

Further, under storm-time conditions, Joule heating and ion drag forcing affect the morphology of the coupled ionosphere-thermosphere system. Upwelling of species (molecules like N${_2}$ and O${_2}$) is caused when the polar upper atmosphere gets heated up. These molecular-rich gases get extended towards the equator due to the ionospheric convection and forces due to the gradient in pressure. This heating effect at the poles induces global wind circulation, causing downwelling of the molecular-rich gases to the lower latitudes, thereby reducing the loss rate of O+ ions in the F-region ionosphere over the low latitudes. This causes a positive storm effect observed as enhancements in the F-region electron density. On the other hand, the negative ionospheric storm phenomenon is observed as reductions or depletions in the electron density (\cite{sc:12,sc:13,sc:14,sc:15,sc:16,sc:17,sc:18,sc:19,sc:20} and references therein). However, any such effects on the ionization over the observed station due to compositional changes can be ruled out from Figure \ref{fig8}. The O/N${_2}$ values over the whole Indian region showed no significant changes that could have affected in terms of enhancements (depletions) in the ionization over the northern crest of EIA as a manifestation of positive (negative) storm effects during October 31 (November 04), 2021. We further add Figure \ref{fig11} showing the O/N${_2}$ map on a quiet day (October 28, 2021) as a reference.

Now in Figure \ref{fig2}, one can observe the geomagnetic storm to be much stronger on November 04 in comparison to that on October 31, 2021. Here we observe a fluctuating IMF $B_z$ between southward and northward orientations as a result of the presence of the sheath region. Although the IMF $B_z$ had been fluctuating during this period, the magnitude of the same had been higher and as a result, we observed growth in the ring current and a strong geomagnetic storm condition prevailed. However, this aspect of a stronger magnitude of the southward IMF component alone was not able to cause any geoeffectivity over the low-latitude ionosphere on that day. 

In contrast, for the much weaker event on October 31, when the magnitude of IMF $B_z$ had been almost half, due to a steady, non-fluctuating IMF $B_z$, we observed significant changes over the low-latitude ionosphere. Hence, it becomes clear in the present study that it is the presence of a non-fluctuating southward $B_z$, for a sufficient interval of time, that was able to create a non-fluctuating or a stable (or DC) prompt penetration electric field that in turn assisted in the strengthening of the low-latitude electrodynamics and hence a greater ionospheric response during the weaker events. This aspect is observed in Figure \ref{fig6}, where during the interval of 08:00-16:00 LT, the residual EEJ had strong positive values that were able to assist in the development of stronger plasma distribution over the crest region. On the other hand, the AC penetration caused as a result of the sheath region and subsequent fluctuating IMF $B_z$ restricted the growth of EEJ strength over these LT hours. Therefore, it is evident that competition between the IMF $B_z$ magnitude and fluctuations (+/-) is something very important for ionospheric responsivity forecasting.    

Further, in a study by \citep{sc:76}, they show that the Disturbance storm time (Dst) or SYM-H index may not be considered a good proxy to estimate the strength and subsequent geoeffectivity. They argued that the Dst index was unable to capture various physical processes in the MI system (particle precipitation, Joule heating, ionospheric outflow, thermospheric density variations, radiation belts' state, etc.) under geomagnetically active conditions. Therefore, in agreement with their study, the SYM-H in the present case was not able to determine the geoeffectivity of the ICME events where IMF $B_z$ polarity had been stable inside the MC-like structure and fluctuating in the sheath region. Under such cases, the penetration electric field (related to the solar wind and the magnetic field dynamics as given in equation \ref{c}) perturbations become an important proxy for the geoeffectivity evaluation (see \cite{sc:71} and references therein). 

\begin{equation}
    E = V_{sw}*B_{IMF}*sin \theta
    \label{c}
\end{equation}

where, $V_{sw}$ is the solar wind velocity, $B_{IMF}$ is the magnetic field strength and $\theta$ is the angle between the IMF and the geomagnetic field.

Furthermore, it is to be noted that magnetic sheaths are, in general, associated with Alfvenic fluctuations in the magnetic field and as a consequence, cause fluctuating PPEF, which can often be responsible for stronger ionospheric response. However, this is not the case in the present study. A few recent studies \citep{sc:81,sc:94} show that a combination of multiple drivers can also generate disproportionate ionospheric impact during geomagnetic storms. Therefore, factors including the background ionospheric condition, meridional wind, etc. can never be neglected and hence are already discussed in this study. It is from this deduction that the major inference drawn in the present work that the non-fluctuating Bz, for a sufficient interval of time, due to the presence of an ICME-driven MC-like region, is responsible for generating stable and consistent penetration effects for the strengthening of the low-latitude electrodynamics, despite the storm being weak, is important and merits attention.

Lastly, identifying such types of events along with the corresponding responses of the ionosphere, as shown in the present study, to perform fruitful statistical inferences, is non-trivial. However, identifying more such cases would be helpful for the space science community to perform a comprehensive and holistic study of the ionospheric responses to varying severity of space weather events in the future. As a follow-up to this work, we will be performing statistical studies as well as investigating the periodicity of the IMF $B_z$ and penetration electric field fluctuations to be able to tell how slow/fast these sheath region fluctuations have to be, to be geoeffective. 

\section{Summary}

This work was directed to reveal that a weaker geomagnetic storm occurring on October 31, 2021, had caused a greater ionospheric response, over the low-latitude region in the Indian sector, compared to a much stronger geomagnetic event of November 04, 2021. The presence of an ICME-driven sheath region, that had caused a fluctuating penetration electric field over the equatorial electrodynamics, was shown to be responsible for the prevention of the development of strong and consistent levels of EEJ strength. This restricted the otherwise expected plasma enhancements over the low-latitude ionosphere during such periods. This investigation brought forward, that for this particular case study and under poleward neutral wind variations, it was neither the storm intensity (SYM-H) nor the IMF $B_z$ magnitude, but the occurrence of the non-fluctuating $B_z$ (for a sufficient interval of time), due to the presence of an ICME-driven MC-like region, responsible for generating stable and consistent penetration effects for the strengthening of the low-latitude electrodynamics, despite the storm being weak. This case study is a path forward for the development of state-of-the-art space weather forecast systems, for reliable predictions related to low- and equatorial electrodynamics under geomagnetically active conditions, by incorporating the time duration for which the IMF remains completely southward without any fluctuations.  

\section*{Acknowledgments}

The authors acknowledge the high-resolution IMF $|B|$, $B_x$, $B_y$, $B_z$, $IEF_y$, SYM-H, Proton Temperature, $V_{sw}$, Proton Density, Flow Pressure, and Plasma $\beta$ in addition to the daily F10.7 data used in this work. Acknowledgments go to the World Data Center of Kyoto University for the $K_p$, and $A_p$ indices. Acknowledgments are due to the Indian Institute of Geomagnetism, Navi Mumbai for providing the EEJ data. We acknowledge the IGS network for providing openly available GPS TEC data for our analysis of ionospheric conditions over Hyderabad and Bengaluru. Further acknowledgments go to the SAC, ISRO for providing a NavIC receiver to PRL, Ahmedabad. Simulation runs obtained using the NCAR-TIEGCM and WACCM-X, hosted by the CCMC, are duly acknowledged. Acknowledgments go to the SME data obtained from the SuperMag network website. The authors thank the PIs of the magnetic observatories and the National institutes that support these observatories. The global thermospheric O/N$_{2}$ maps obtained from the GUVI onboard the TIMED spacecraft hosted by the JHUAPL are acknowledged. We thank both reviewers for their valuable comments and suggestions that have enhanced the quality of the manuscript. This work is supported by the Department of Space, Government of India.

\bibliographystyle{jasr-model5-names}
\biboptions{authoryear}
\bibliography{refs}

\end{document}